\documentclass[
 aip,
 amsmath,amssymb,
 reprint,%
]{revtex4-2}

\usepackage{graphicx}
\usepackage{dcolumn}
\usepackage{bm}

\usepackage[utf8]{inputenc}
\usepackage[T1]{fontenc}
\usepackage{mathptmx}
\usepackage{etoolbox}
\usepackage{subcaption}
\usepackage{siunitx}
\usepackage{pgfplots}
\usepackage{hyperref}
\usepackage{tikz}
\usepackage{pdfpages}
\pgfplotsset{compat=1.18}
\usepackage{ragged2e}
\usepackage[justification=justified,singlelinecheck=false]{caption}
\usepackage{natbib}
\usepackage{hyperref}
\hypersetup{
    colorlinks=true,
    linkcolor=blue,
    citecolor=blue,
    urlcolor=blue
}

\makeatletter
\def\@email#1#2{%
 \endgroup
 \patchcmd{\titleblock@produce}
  {\frontmatter@RRAPformat}
  {\frontmatter@RRAPformat{\produce@RRAP{*#1\href{mailto:#2}{#2}}}\frontmatter@RRAPformat}
  {}{}
}%
\makeatother

\makeatletter
\AtBeginDocument{\let\LS@rot\@undefined}
\makeatother

\begin{document}

\title{\huge Band Meandering due to Charged Impurity Effects and Carrier Transport in Ternary Topological Insulators}

\author{Kanav Sharma}
\affiliation{Department of Physical Sciences, Indian Institute of Science Education and Research Kolkata, Nadia, 741246, West Bengal, India}

\author{Niranjay K R}
\affiliation{Department of Physical Sciences, Indian Institute of Science Education and Research Kolkata, Nadia, 741246, West Bengal, India}

\author{Infan S Mesh}
\affiliation{Department of Physical Sciences, Indian Institute of Science Education and Research Kolkata, Nadia, 741246, West Bengal, India}

\author{Radha Krishna Gopal}
\affiliation{Department of Physics and Material Sciences and Engineering, Jaypee Institute of Information Technology, Sector 62, Noida, India}

\author{Chiranjib Mitra}
\affiliation{Department of Physical Sciences, Indian Institute of Science Education and Research Kolkata, Nadia, 741246, West Bengal, India}
\email{chiranjib@iiserkol.ac.in}

\date{\today}

\begin{abstract}
\large
\raggedright
Controlling charged impurity disorder is a critical challenge for realizing the promise of topological insulator (TI) surfaces in devices. While doping is often used to tune the chemical potential, its impact on the fundamental disorder landscape remains poorly understood. Here, we investigate this effect in ternary (Bi,Sb)$_2$Te$_3$ (BST) thin films and their indium-doped (IBST) counterparts. Gate-dependent transport reveals that indium doping increases charged impurity density by an order of magnitude, which in turn reduces the characteristic size of disorder-induced charge puddles from $\sim$91 nm to $\sim$38 nm. This amplified disorder enhances Coulomb scattering and suppresses field-effect mobility, directly demonstrating how doping-induced compensation degrades surface transport. Our work establishes doping as a powerful method to probe the limits of topological protection and underscores that defect suppression, not just compensation, is essential for developing high-performance TI devices.
 
\end{abstract}

\maketitle
\large
Dirac materials\cite{colloquium,Kane2005,t3,t4} having insulating bulk with maximum conduction from the surface states are highly desirable because of its unique properties like spin-momentum locking, which suppresses backscattering from non-magnetic impurities and promises applications in spintronics, quantum computation, and high-speed electronics .~\cite{spintronics,pal2024enhancement,superconducting} However, realizing these applications requires effective control of the chemical potential, ideally tuned to the Dirac point, where intrinsic transport properties are most robust.

A fundamental challenge in realizing pristine topological surface states is the ubiquitous disorder from charged impurities (donors and acceptors), which creates strong spatial fluctuations in the Coulomb potential. In the bulk of these narrow-gap semiconductors, these fluctuations can be of the order of the band gap itself, causing extreme band bending and the formation of bulk electron-hole puddles via a nonlinear screening process \cite{band_meandering,shklovskii2013electronic}. This bulk disorder also manifests on the surface: while the metallic topological surface state provides a strong screening channel, its effectiveness is limited near the charge neutrality point. The unscreened residual potential fluctuations rigidly shift the local Dirac cone, creating nanoscale electron and hole puddles. Recent studies on compensated systems like BiSbTeSe$_2$ reveal that these surface puddles exhibit a characteristic length scale of $40-50$ nm. Crucially, the persistence of these puddles indicates that screening is not solely a surface effect but likely a cooperative process between the surface state and the bulk puddles themselves \cite{Knispel}. These inhomogeneities act as scattering centers, degrading carrier mobility and obscuring the intrinsic Dirac fermion transport, a phenomenon extensively studied in graphene \cite{backscttering} and now recognized as a critical factor in three-dimensional topological insulators.

While previous studies have highlighted the importance of charge puddles in compensated topological insulators, several key aspects remain unresolved. Borgwardt \textit{et al.}\cite{Borgwardt} demonstrated the self-organization and temperature-driven evaporation of bulk puddles in BiSbTeSe$_2$ using optical spectroscopy, but their work was restricted to thick single crystals and explicitly excluded surface contributions. Knispel \textit{et al.}\cite{Knispel} extended this approach by combining optical and STM/STS measurements on BiSbTeSe$_2$ to probe both bulk and surface puddles. However, their study did not clarify how puddles influence charge transport. Moreover, neither work addressed ternary thin films, intentional doping, or substrate/interface disorder, all of which are directly relevant for device performance. 

In this work, we address these open issues by using intentional indium doping as a controlled method to amplify the charged impurity landscape in ternary (Bi,Sb)$_2$Te$_3$ (BST) thin films. We demonstrate that increased doping density leads directly to a reduction in the characteristic surface puddle size and a suppression of carrier mobility. This approach allows us to quantitatively establish the relationship between impurity concentration, puddle formation, and transport degradation. Furthermore, we observe a temperature-driven n-p transition linked to these disorder-induced p-n regions. By systematically varying the disorder through doping, our study reveals that mitigating intrinsic defect concentrations, rather than merely compensating them, is the fundamental requirement for achieving high-mobility topological surface states in devices.

Pulsed laser deposition (PLD)\cite{pulsed,pld2,mitra2002magnetotransport} is a versatile, cost-effective technique for growing polycrystalline crystals to study quantum properties of materials. It uses high-energy laser pulses to ablate a target, depositing atoms and ions onto a substrate to form thin films. PLD enables precise control of stoichiometry and thickness, making it ideal for fabricating high-quality topological insulators, complex oxides, and superconductors. By adjusting laser energy, substrate temperature, and gas pressure, researchers can tailor film characteristics. We utilized this established method for consistent material growth, ensuring precision in our process. All details of characterisation are given elsewhere.\cite{sharma2024suppression}

\begin{figure}[t]
    \centering
    \includegraphics[width=0.5\textwidth]{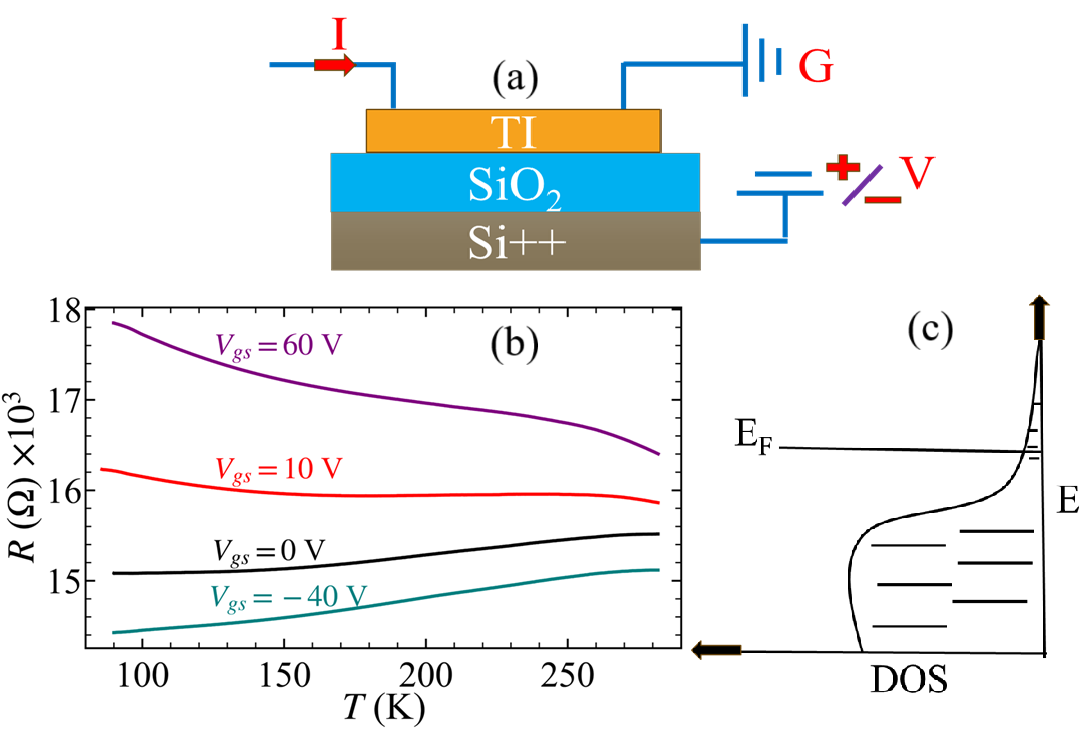}
     \vspace{0.05pt}
    \caption{\justifying (a) illustrates a thin film of topological insulator placed on a Si/SiO2 substrate, with a positive or negative gate voltage (+/- V) applied to the Si++ wafer. The current through the topological insulator is denoted as I, and G represents the common ground. (b) Shows resistance vs temperature at different gate voltages denoted by $V_{gs}$ in volts. (c) Shows the density of states (DOS) of the valence band versus energy (E) with chemical potential $E_F$ at the band tail.}
    \label{fig:1}
    \vspace{6pt} 
\end{figure}

\begin{figure}[!htbp]
    \centering
    \includegraphics[width=0.5\textwidth]{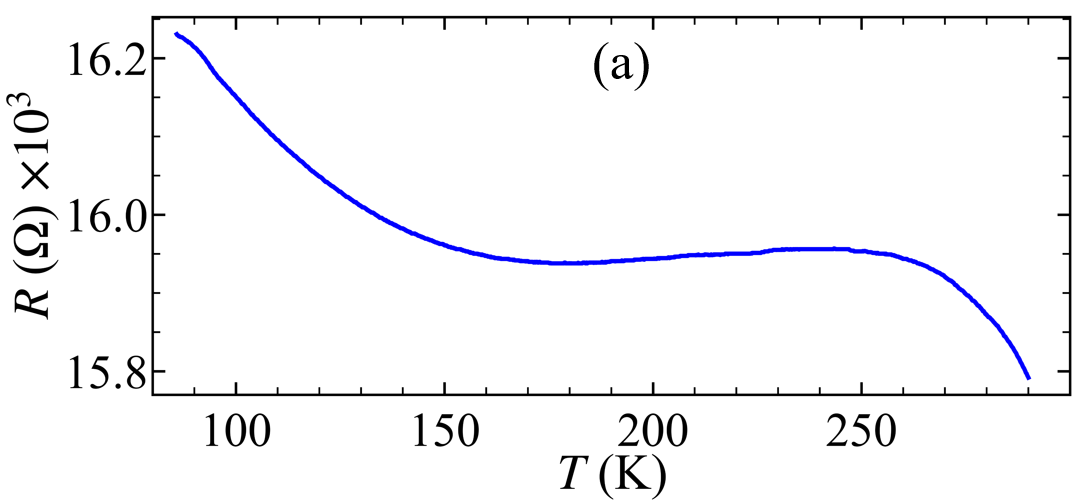}\hfill
    \includegraphics[width=0.5\textwidth]{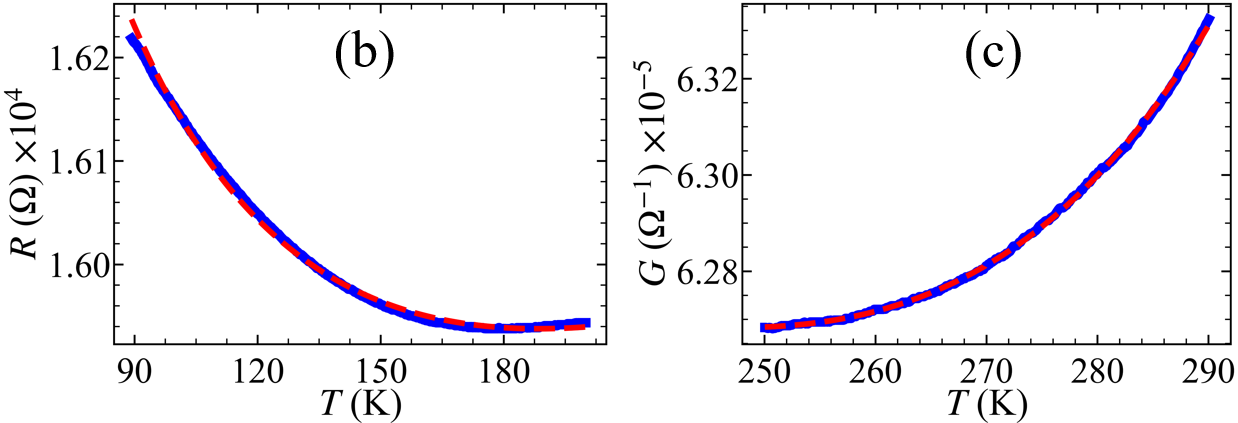}
    \includegraphics[width=0.5\textwidth]{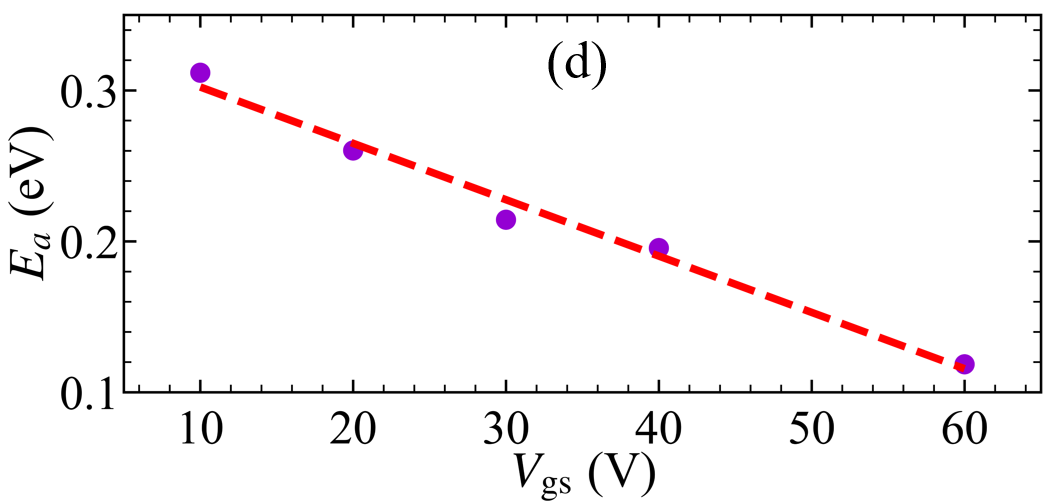}
    
    \vspace{4pt}
    \caption{\justifying 
    (a) presents the resistance–temperature (R–T) behavior at $V_{gs} = 10$ volts, fitted using two different approaches depicted in (b) and (c). 
    (b) applies the Mott variable range hopping model, utilizing Equation 1 for fitting. 
    (c) employs the parallel resistor model, based on Equation 2. 
    (d) displays activation energies at elevated temperatures for various gate voltages, where their slopes can be utilized to determine the density of states around $E_F$.
    }
    \label{fig:2}
\end{figure}

The chemical potential can be modified either by applying gate voltage or by doping.\cite{gate1,gate2,gate3,dope1,dope2,dope3} Adjusting the chemical potential provides valuable insights into material properties, particularly in small band gap materials. We have grown three materials: BST and IBST with a thickness of 30 nm, which would be referred to as thin films, and BST with a thickness of 60 nm, which would be referred to as thick films. Fig.~\ref{fig:1}(a) illustrates the specific device used for gating, featuring a 300 nm thick $SiO_{2}$ layer. The BST thin film shows metallic behaviour for <= 0  gate bias and insulating behaviour for >0 gate bias as shown in Fig.~\ref{fig:1}(b) which corresponds to p-type behaviour. In the insulating regime at low temperatures, it displays Mott variable range hopping. This shows the chemical potential to be inside the valence band tail, as depicted in Fig.~\ref{fig:1} (c) showing the density of states versus energy plot. Fig.~\ref{fig:2}(a) illustrates the resistance-temperature behavior at a gate voltage of 10 volts. Fig.~\ref{fig:2}(b) displays the lower temperature region (less than 200K), which was fitted using Mott variable-range hopping given by\cite{mott1}$^,$\cite{mott2} 
\begin{equation} R(T) = A \cdot T^{0.25} \cdot \exp{\left(\left(\frac{T_o}{T}\right)^{0.25}\right)}
\end{equation}   where To is the characterstic temperature. The higher temperature part was fitted using the parallel resistor model given by\cite{singh2017linear}$^,$\cite{mitra2001growth}
\begin{equation} 
G(T) = \frac{1}{a + bT} + \frac{1}{c \, \exp\left(\frac{\Delta E}{K_b T}\right)}
\end{equation} 
yielding an activation energy $(\Delta E)$ of 312 meV.

The insulating behavior observed at elevated temperatures across all positive gate voltages enables us to calculate the activation energy. At $V_{gs}$ = 10 volts, the fit to the activated part of equation 2 is displayed by the red curve in Fig.~\ref{fig:2}(c). From the resistance versus temperature curve (Fig.~\ref{fig:1}(b)) it is evident that the activation energy can be determined only for positive gate voltage data. The extracted activation energy values are plotted against gate voltages in Fig.~\ref{fig:2}(d). The slope of this plot enables us to estimate the density of states near the chemical potential using quantum capacitance formulations given by,\cite{quantum_capacitance}  
\begin{equation} 
\frac{dE_F}{dV_g} = -\frac{dE_a}{dV_g} = \frac{eC_{ox}}{C_{ox} + C_d}
\end{equation}
where $C_d = e^{2}D(E)$ and $C_{ox}$ is the oxide capacitance per unit area. The density of states D(E) at the chemical potential was extracted which came out to be $1.3 \times 10^{13} \text{eV}^{-1} \text{cm}^{-2}$.  

We have also calculated the four-probe mobility given by,\cite{mobility_formula} 
\begin{equation} \mu = \left(\frac{L}{W \, C_{\text{ox}} \, V_{\text{ds}}}\right) \cdot \frac{\partial I_{\text{ds}}}{\partial V_{\text{gs}}}
\end{equation}
\begin{figure}[t]
    \centering

    \includegraphics[width=0.5\textwidth]{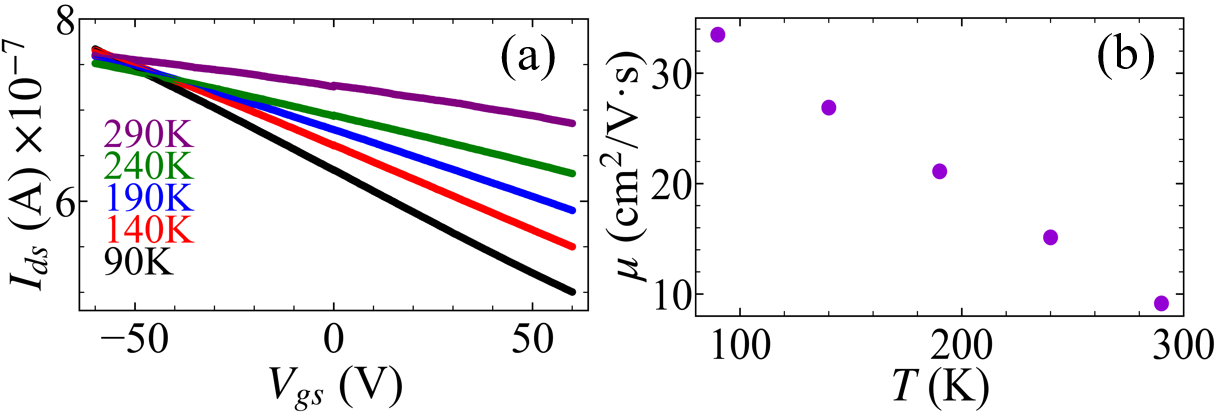} \hfill
    \includegraphics[width=0.5\textwidth]{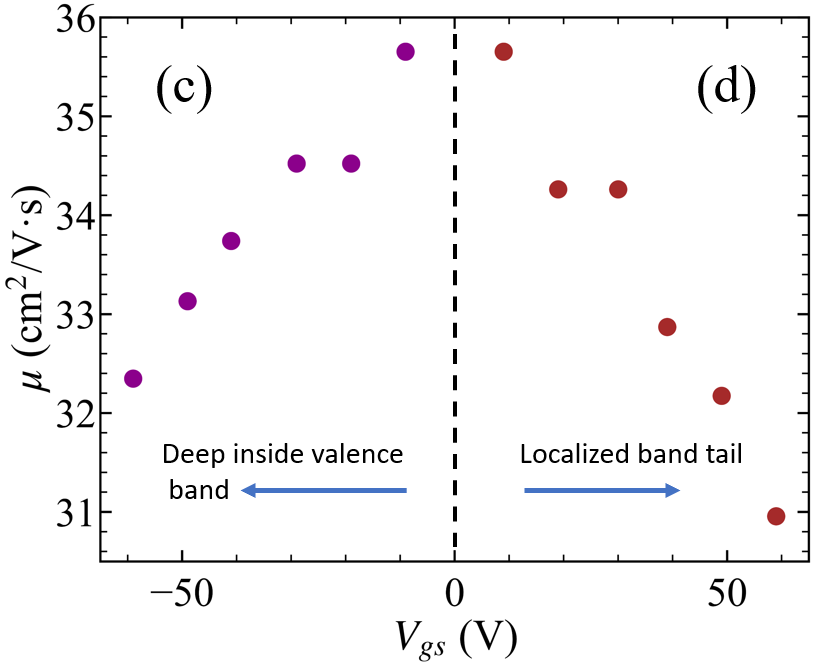}
    
    \caption{\justifying (a) depicts the relationship between drain-source current and gate voltage at various temperatures.
(b) presents the field-effect mobility as a function of temperature, derived using Equation 3. (c,d) shows the decreasing behaviour of mobility while going to either side of the band.
}
\label{fig:3}
\end{figure}

\begin{figure*}[t] 
    
    \includegraphics[width=1.0\textwidth]{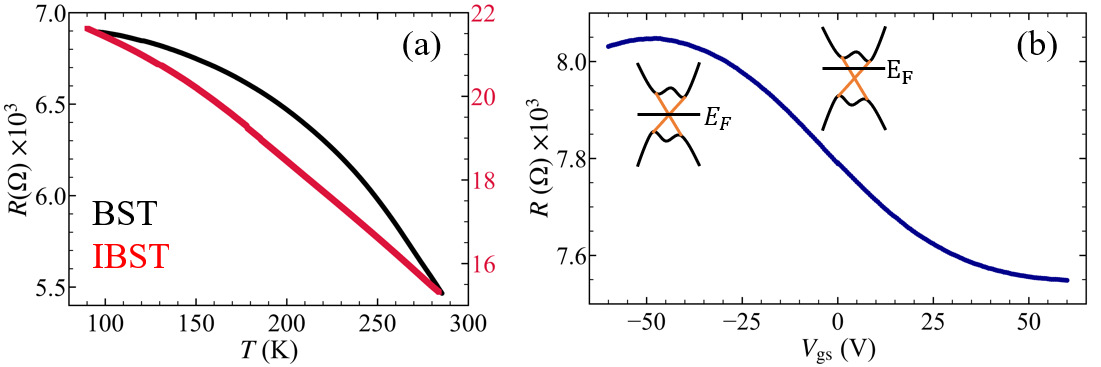}

    \includegraphics[width=1.0\textwidth]{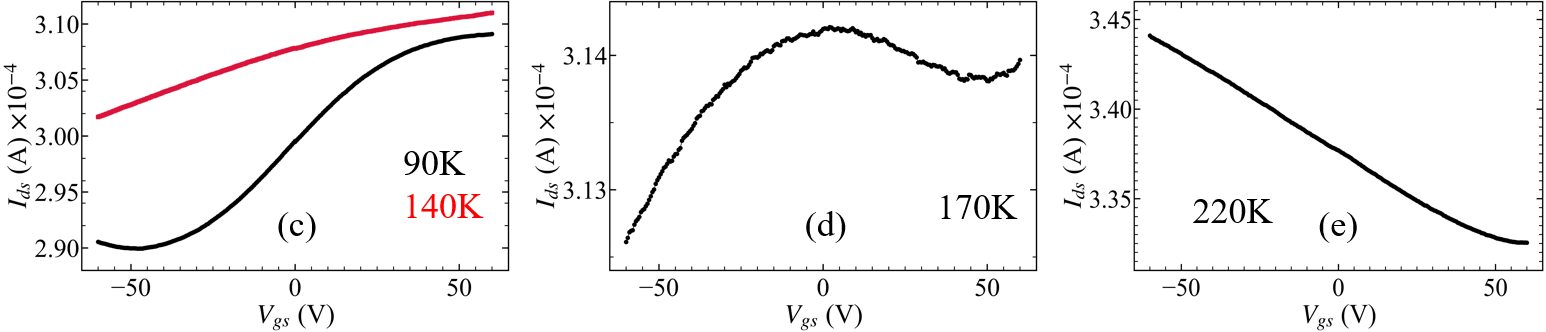}
    \includegraphics[width=0.5\textwidth]{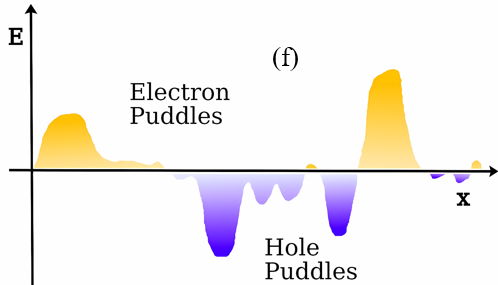}
     \vspace{20pt}
    \caption{\justifying (a) Shows R-T of thick-BST and thin-IBST. (b) The R-Vgs plot demonstrates a peak in resistance, corresponding to the charge neutrality point for BST. 
    (c,d,e) depicts the relationship between drain-source current and gate voltage at various temperatures for BST. (f) depicts the electron-hole puddles.}
    \label{fig:4}
     \vspace{1pt}
\end{figure*} 
 At different temperatures, the drain-source current converges at -60 V, indicating that a negative voltage shifts the chemical potential further inside the valence band, as depicted in Fig.~\ref{fig:3}(a). Fig.~\ref{fig:3}(b) demonstrates that the mobility increases with decreasing temperature, which is attributed to the dominance of surface states at lower temperatures. This observed order of mobility is consistent with recent findings based on Hall measurements.\cite{mobility_pld} However, the mobility values remain relatively low, likely due to the polycrystalline nature of the PLD-grown samples, which include grain boundaries and exhibit a more disordered structure, influencing the observed behavior.From the gate-voltage dependence of mobility at 90~K shown in Fig.~\ref{fig:3}(c,d), it is evident that the mobility decreases as the chemical potential moves either toward localized states near the band tail or deep into the valence band, where enhanced scattering suppresses carrier transport. Further we have calculated charged impurity density $n_{imp}$ around 90K from D(E) in the thermal window $K_{B}*T$ gives $n_{imp} = 10^{11}cm^{-2}$. This impurity density could be the effect of oxide traps of silicon dioxide as it does not matches with the intrinsic doping value n.\cite{burson2013direct}$^,$\cite{PhysRevB.79.205411} However, no quantitative estimate is available to prove this possibility.

The temperature dependence of resistance for thick BST and thin IBST is shown in Fig.~\ref{fig:4}(a). The activation energies for BST and IBST, calculated using Equation 2, are 114 meV and 65.7 meV, respectively. The insulating behavior indicates that the chemical potential lies within the bulk band gap, where the density of states is low, and local impurities may cause potential fluctuations.
In the case of BST, at 90K, the resistance displays a broad hump as a function of gate voltage between -40 to -58 Volts, as shown in Fig.~\ref{fig:4}(b). The span of this hump occurs over a broad range of gate voltage, likely due to CNP smearing within our material. The schematic given in the inset of Fig.~\ref{fig:4}(b) provides a depiction of the chemical potential's position inside the band for two different gate voltages. At CNP, the intrinsic doping is defined as\cite{gate1}:
\begin{equation}
\Delta n = \frac{C_{\text{ox}} V_{\text{CNP}}}{e}
\end{equation}

The calculated carrier density \( \Delta n \) is \( 3.95 \times 10^{12} \) cm\(^{-2}\). Our result aligns well with findings from similar thin-film TI studies reported in the literature \cite{Rosen}.

Next, drawing from Equation 1a in the work by Kim et al. \cite{charged_imprity_nature}, we computed the density of charged impurities, \( n_{\text{imp}} \). This value turned out to be \( 2.23 \times 10^{14} \) cm\(^{-2}\), using a constant \( C = 30 \) (an empirical factor that accounts for the strength of impurity scattering in the model). Such a high impurity density indicates that these charged impurities are not extrinsic  but rather intrinsic \cite{charged_impurity}.

\begin{table*}[t]
\centering
\renewcommand{\arraystretch}{1.3}

\begin{tabular}{|c|c|c|c|}
\hline\hline
\textbf{Sample name} &
\textbf{Ref.} &
\textbf{Amplitude of fluctuations $\Gamma$ (meV)} &
\textbf{Characteristic size of puddles $r_s$ (nm)} \\
\hline
Doped $Bi_{2}Se_{3}/Bi_{2}Te_{3}$ &
H. Beidenkopf et al.\cite{beidenkopf2011spatial} &
10--20 &
20--30 \\
\hline
--do-- &
B. Skinner et al.\cite{skinner2013theory} &
$\sim$18 &
$\sim$5 \\
\hline
$BiSbTeSe_{2}$ &
T. Knispel et al.\cite{Knispel} &
8--14 &
40--50 \\
\hline
--do-- &
Yi Huang et al.\cite{huang2021disorder} &
$\sim$30-35 &
$\sim$5-10 \\
\hline
BST (60 nm) &
Our work &
$\sim$4.86 &
$\sim$91 \\
\hline
IBST (60 nm) &
Our work &
-- &
$\sim$38 \\
\hline\hline
\end{tabular}

\vspace{0.1cm}
\caption{Amplitude of potential fluctuations and characteristic surface puddle sizes reported for various topological insulator systems.}
\label{tab:puddles}
\end{table*}

To further validate this, we employed the following equation for field-effect mobility (\( \mu_{\text{FE}} \)), which describes how efficiently charge carriers move under an applied electric field in a field-effect transistor setup \cite{charged_imprity_nature}:

\begin{equation}
\mu_{\text{FE}} = \frac{C \cdot e}{n_{\text{imp}} \cdot h}
\end{equation}

\begin{figure}[t]
    \centering
    \includegraphics[width=0.5\textwidth]{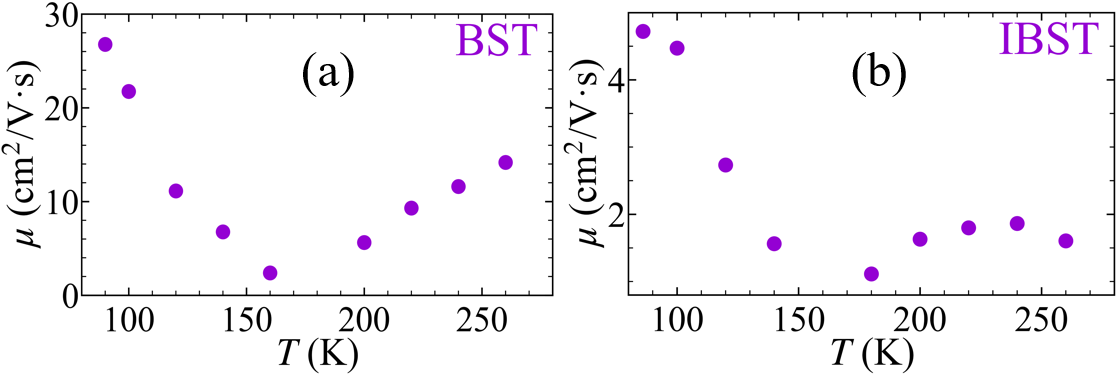}
     \vspace{0.5pt}
    \caption{\justifying (a,b) illustrates the temperature variation of the magnitude of four-probe field-effect mobility for BST and IBST respectively.}
    \label{fig:5}
\end{figure}

Here, \( e \) is the elementary charge, and \( h \) is Planck's constant. Plugging in the value of \( n_{\text{imp}} \) from above yields \( \mu_{\text{FE}} = 32.5 \) cm²/V·s. This calculated mobility is quite close to the value we derived using Equation 4 at a temperature of 90 K, as illustrated in Fig.~\ref{fig:5}(a). The agreement between these two approaches confirms that our choice of \( C = 30 \) is reasonable and appropriate for this system, as it bridges theoretical modeling with experimental observations.

Fig.~\ref{fig:4}(c) displays the drain-source current (\( I_{\text{ds}} \)) as a function of gate voltage (\( V_g \)), revealing the material's n-type behavior at low temperatures. However, an intriguing slope change of $I_{ds}-V_{g}$ occurs at an intermediate temperature around 170 K, as shown in Fig.~\ref{fig:4}(d). At higher temperatures, the material exhibits p-type characteristics, where holes dominate, as depicted in Fig.~\ref{fig:4}(e) \cite{pn_transition, pn_transition2}.

This temperature-driven shift from n-type to p-type conduction can be attributed to the formation of p-n regions—localized areas where electron-rich (n-type) and hole-rich (p-type) zones coexist—near the chemical potential. These regions stem from potential fluctuations, which are variations in the local electrostatic potential caused by charged impurities. Such impurities could be intrinsic to the topological insulator material or introduced extrinsically via oxide traps in the underlying \( \mathrm{SiO}_2 \) substrate (a common dielectric layer in device fabrication that can trap charges at interfaces). At low temperatures, the chemical potential lies close to the conduction band, promoting electron-dominated conduction primarily through percolating electron-rich puddles. As the temperature rises to approximately 170 K, thermal energy becomes sufficient to activate mobile holes in the valence band tails of the hole-rich puddles. This activation provides a percolation path for hole conduction, which begins to compete with and eventually surpass electron conduction. This crossover, driven by the thermally-enabled mobilization of holes within the disorder-induced landscape, leads to a dominance of p-type behavior.

To quantify the disorder further, we calculated the Coulomb interaction energy between neighboring dopants (impurities) using\cite{dopant_interaction}:

\begin{equation}
E_C = \frac{e^2 \cdot N_{\text{imp}}^{1/3}}{4\pi \epsilon}
\end{equation}

Where $N_{imp} = \frac{n_{imp}}{t}$, t is the thickness. With a dielectric constant \( \epsilon = 200 \epsilon_0 \) (where \( \epsilon_0 \) is the vacuum permittivity, and the factor of 200 reflects the high permittivity typical in such materials), we obtained \( E_C = 2.4 \) meV. Building on insights from Skinner et al., the characteristic size of surface "puddles"—inhomogeneous charge accumulations forming electron or hole lakes due to potential fluctuations—depends on whether the chemical potential \( E_F \) is close to the Dirac point or farther away. To assess this, we first computed the proximity of \( E_F \) from the Dirac point (or the relevant energy shift) as\cite{shyaga2025multi}:

\begin{equation}
\Delta E_F = \frac{\pi \hbar^2 \Delta n}{2m^*}
\end{equation}

Using an effective mass \( m^* = 0.32 m_e \) (where \( m_e \) is the electron rest mass, and \( \Delta n \) represents the change in carrier density), we found \( \Delta E_F  = 14.75 \) meV. This relatively small value suggests \( \Delta E_F  \) is sufficiently close to the Dirac point, allowing us to apply the formula for the puddle radius \( r_s = \left( 4 \alpha_{\text{eff}}^4 N_{\text{def}} \right)^{-1/3} \), where \( \alpha_{\text{eff}} \) is the effective fine-structure constant (accounting for dielectric screening) and \( N_{\text{def}} \) is the defect density. The resulting \( r_s = 91 \) nm matches well with both experimental observations and theoretical predictions from prior studies.

Additionally, the disorder parameter $\Gamma=\frac{4E_{C}}{\alpha_{eff}^{2/3}}$ evaluates to 66.7~meV. This theoretical value is an order of magnitude larger than our experimentally estimated value of $\Gamma \approx 4.86$~meV, which we calculated from the CNP smearing in the -40 to -58~V region [Fig.~\ref{fig:4}(b)]. This large discrepancy provides strong experimental confirmation of the findings by Knispel \textit{et al.}. Our results support their conclusion that the simple Thomas-Fermi screening model is insufficient to describe puddle amplitude, likely due to a cooperative screening effect between bulk and surface carriers. 

We hypothesize that the temperature-dependent transition from n-type to p-type at around 170 K in Fig.~\ref{fig:4}(d), mirrors the transition seen at a fixed 90 K, which was obtained around -55 Volts shown in Fig.~\ref{fig:4}(b), when we varied the gate voltage. The corresponding energy shift \( \Delta E_F / k_B \) (where \( \Delta E_F \) is the Fermi level shift and \( k_B \) is Boltzmann's constant) equates to 171.1 K, closely matching the observed transition temperature. 
We note that this 171~K transition ($\approx 14.75$~meV) represents the activation energy for hole percolation through the surface puddle landscape. This \textit{surface} transport barrier is a distinct energy scale from the \textit{bulk} Coulomb interaction $E_C$ (2.4~meV $\approx$ 30~K) calculated earlier. This distinction is consistent with studies like Knispel \textit{et al.}, which also observed that surface puddles persist at temperatures well above the bulk puddle evaporation scale.
This equivalence suggests that thermal effects and electrostatic gating induce similar modulations of the chemical potential.

The experimentally determined puddle size provides a baseline for understanding the intrinsic disorder in our BST films. We now use indium doping as a tool to deliberately modify this disorder landscape. If the theory holds that puddle size is governed by impurity density, we expect increased doping to result in a higher $n_{\text{imp}}$ and consequently a smaller $r_s$. Indium doping introduces compensation—balancing donor and acceptor impurities—which further shifts the chemical potential in the thin film. In IBST, the charged impurity density \( n_{\text{imp}} \) surges to \( 1.53 \times 10^{15} \) cm\(^{-2}\), an order of magnitude higher than in undoped BST. This elevated density confirms that impurities are primarily intrinsic and are amplified by the indium incorporation, exacerbating potential fluctuations and promoting more extensive band meandering and p-n region formation, as described earlier.

The Coulomb interaction in IBST, recalculated using the same formula, yields \( E_C = 5.7 \) meV, demonstrating that increased compensation enhances electrostatic repulsion between dopants. Consequently, the characteristic surface puddle size shrinks to \( r_s = 38 \) nm, and the disorder parameter \( \Gamma \) rises to an overestimated 158.3 meV (again, due to the model's limitations in handling bulk-surface interactions). Table I presents the values of  $r_s$ and $\Gamma$ reported in different studies.
The key insight from our doping experiment is clear: increasing the charged impurity density through indium doping intensifies Coulomb interactions, which directly shrinks the characteristic puddle size from 91 nm to 38 nm. This quantitative result provides compelling experimental evidence for the theoretical relationship between impurity density and puddle size. These smaller, more numerous puddles create a stronger scattering landscape for surface electrons, thereby drastically diminishing the field-effect mobility in IBST compared to BST (Fig.~\ref{fig:5}(b)). This demonstrates conclusively that impurity-driven scattering can override the topological protection of surface states. The counter-intuitive outcome—that doping intended to tune the Fermi level instead degrades performance—establishes that meticulous suppression of intrinsic impurity concentrations, not just compensation, is the essential path toward high-mobility topological devices.

In conclusion, we have used indium doping as a strategic tool to amplify and quantify the role of charged impurities in topological insulators. We demonstrated that increased impurity density shrinks the characteristic charge puddle size from $\sim$91~nm to $\sim$38~nm and suppresses mobility, providing a direct experimental verification of the disorder mechanisms that degrade surface transport. The observed temperature-driven n-to-p transition at $\sim$170~K further reflects the complex electrostatics of this disorder landscape. Our findings pivot the development strategy for TI-based devices: they prove that defect suppression at the source, rather than post-growth compensation, is the critical requirement for preserving the topological surface state. Future work must focus on synthetic techniques that minimize intrinsic impurity concentrations to unlock the full potential of topological insulators for quantum technologies.\\

The authors acknowledge the University Grant Commission (UGC) of the Government of India for financial support.

\begin{flushleft}
\textbf{\normalsize{AUTHOR DECLARATIONS}}\\
\textbf{\normalsize{Conflict of Interest}}\\
\hspace{15pt}The authors have no conflicts to disclose.
\end{flushleft}

\begin{flushleft}
\textbf{Author Contributions}
\vspace{-9pt}
\begin{justify}
\textbf{\normalsize{Kanav Sharma}}: Conceptualization (lead); Formal analysis (lead); Data curation (lead); Investigation (lead); Writing – original draft (lead);  
\textbf{\normalsize{Niranjay K R}}: Data curation (supporting); Software (supporting);  
\textbf{\normalsize{Radha Krishna Gopal}}: Review \& editing (supporting).  
\textbf{\normalsize{Chiranjib Mitra}}: Formal analysis (equal); Supervision (lead); Writing – review \& editing (equal).
\end{justify}
\end{flushleft}

\begin{flushleft}
\textbf{\normalsize{DATA AVAILABILITY}}

\begin{justify}
{\vspace{-10pt}
The data that support the findings of this study are available from the corresponding author upon reasonable request.}
\end{justify}
\end{flushleft}

\begin{flushleft}
\textbf{\normalsize{REFERENCES}}
\vspace{-1.1em}
\end{flushleft}

\bibliography{aipsamp}

\end{document}